\documentclass[aps,prc,superscriptaddress,nofootinbib,showpacs,noshowkeys,twocolumn,amssymb,amsmath]{revtex4}
\usepackage[dvips]{graphics,graphicx}
\def\beq{\begin{equation}}
\def\enq{\end{equation}}
\def\beqa{\begin{eqnarray}}
\def\enqa{\end{eqnarray}}
\def\nnb{\nonumber}
\begin{document}
\title{Width of exotics from QCD sum rules : tetraquarks or molecules?}
\author{Su Houng Lee}
\email{suhoung@phya.yonsei.ac.kr}
\author{Kenji Morita}
\email{morita@phya.yonsei.ac.kr}
\affiliation{Institute of Physics and Applied Physics, Yonsei
University, Seoul 120-749, Korea}
\author{Marina Nielsen}
\email{mnielsen@if.usp.br}
\affiliation{Instituto de F\'{\i}sica, Universidade de S\~{a}o Paulo, C.P. 66318, 05389-970 S\~{a}o Paulo, SP, Brazil}
\date{\today}
\begin{abstract}
 We investigate the widths of the recently observed charmonium like resonances
$X(3872), Z(4430)$ and $Z_2(4250)$ using QCD sum
 rules. Extending previous analyses regarding these states as
 diquark-antiquark states or molecules of $D$ mesons, we introduce the
 Breit-Wigner function in the pole term.
 We find that introducing the width increases the mass at small Borel window
 region. Using the operator-product expansion up to dimension 8, we find
 that the sum rules based on interpolating current with molecular
 components give a stable Borel curve from which both the masses and widths of these resonances can be
 well obtained.  Thus the QCD sum rule approach strongly favors the molecular description of these
 states.
\end{abstract}

\maketitle
\section{Introduction}
Since the discovery of a charmonium like resonance $X(3872)$ by Belle
Collaboration \cite{choi03_belle_x3872}, plenty of similar resonances have
been observed in the decay of $B$ mesons. In particular, the recently observed
$Z^+(4430)$ in $\pi^+ \psi'$ invariant mass spectrum has a charge
\cite{choi08_belle_Z4430}, and consequently  can not be a simple charmonium. There are already a number of interpretations on
the structure of these resonances \cite{review}. Although the nature of these states is still an 
open question, tetraquark state and molecular state are both
intriguing possibilities. Since the decay products are a charmonium and a
pion, it is natural to expect that the parent contains four
quarks including $c$ and $\bar{c}$. On the other hand, the masses of
$X(3872)$, $Z^+(4430)$ and the recently observed resonancelike structure
$Z_2^+(4250)$ \cite{mizuk_belle_z1z2},
are very close to thresholds of
two $D$-meson states, $D^* D$, $D^* D_1$ and $D_1 D$, respectively, and,
therefore, it is very tempting to interpret these states as molecular states.

Motivated by these facts, QCD sum rules (QCDSR) have been
extensively used to study these resonances.
In Ref.~\cite{matheus07}, $X(3872)$ was analyzed by assuming it to be a $J^{PC}=
1^{++}$ tetraquark ($c\bar{c}q\bar{q}$) state. Although the result
agreed with the experimental data, an analysis using a current composed
of a $D^* D$ molecule shows better operator-product expansion (OPE) convergence and closer agreement with experimentally observed mass \cite{lee_dsdstar}, strongly suggesting a molecular nature of  $X(3872)$.  

In Ref.~\cite{Lee_Mihara_Navarra_Z4430_PLB}, $Z^+(4430)$ was considered as a
$D^* D_1$ molecule and a good agreement  with data was obtained, while
the tetraquark description has been found to be unsatisfactory
\cite{bracco_z4430tetra}.
Similarly, we have applied the molecular description to the most recent data
of $Z_1^+(4050)$ and $Z_2^+(4250)$ \cite{mizuk_belle_z1z2}, and found that a
$D_1 D$ molecular state can be attributed to $Z_2^+(4250)$. However, it was
not possible to explain $Z_1^+(4050)$ as a molecular $D^* D^*$ state
\cite{lee_morita_nielsen_molecule}.
In all of the calculations above, however, small but finite width was not
taken into account, in spite of that fact that $Z^+(4430)$ and
$Z_2^+(4250)$ have  widths
$\Gamma_{Z^+(4430)} = 45^{+18+30}_{-13-13}$ MeV~\cite{choi08_belle_Z4430} and
$\Gamma_{Z_2^+(4250)} = 177^{+54+316}_{-39-61}$
MeV~\cite{mizuk_belle_z1z2}, respectively.
Of course these widths are much smaller than their masses, around $4$ GeV.
However, the effect of the width should be examined in order to clarify the
structure of these states.  

In this paper, we extend the previous analyses
\cite{matheus07,Lee_Mihara_Navarra_Z4430_PLB,lee_dsdstar,bracco_z4430tetra,lee_morita_nielsen_molecule}
to include the effect of finite width and give further consideration on
the possibility that these states can be considered as tetraquarks or
molecules. The width is usually not calculated in QCD sum rule approaches as the OPE are usually restricted to three terms; perturbative, dimension four and dimension six terms.  Hence the phenomenological sides are composed of three unknown parameters; mass, continuum, and overlap constant.   In the present analysis, the OPE are composed of operators with four different dimensions.  Therefore, an analysis including the width is sensible.

In the next section, we give a brief review of our QCDSR analyses.
In Sec.~\ref{sec:width_general}, we discuss some general features
of effect of finite width. Quantitative analyses of the exotic states are
given in Sec.~\ref{sec:results}. Section \ref{sec:summary} is devoted
to the summary.

\section{QCD sum rules}

The QCD sum rules for mesons are based on the two-point function of a
current $j(x)$ describing a desired state
\begin{equation}
 \Pi(q) = i \int d^4 x e^{iq\cdot x} \langle 0 | T[j(x)j^\dagger(0)]
  |0\rangle,
\end{equation}
and the dispersion relation
\begin{equation}
 \Pi(q) = \int ds \frac{\rho(s)}{s-q^2} + (\text{subtraction terms})
\end{equation}
with $\rho(s)$ being the spectral density.
While computing the two-point function in terms of quarks and gluons by
making use of operator product expansion (OPE), which takes into account
non-perturbative effect through QCD condensates, one models the hadronic
spectral density $\rho^{\text{phen}}(s)$ with a pole describing the
ground state and a continuum,
namely,
\begin{equation}
 \rho^{\text{phen}}(s) = \rho^{\text{pole}}(s) + \rho^{\text{cont}}(s).
\end{equation}
In the narrow width approximation, the pole part of the hadronic
spectral density is set to a delta function
$\rho^{\text{pole}}(s) = \lambda^2 \delta(s-m^2)$,  with $\langle 0 | j |\text{meson} \rangle = \lambda$ being the overlap of the current and the physical meson.
In this work,  we replace this part with the relativistic
Breit-Wigner function to take the width into account. The continuum
part above the threshold $s_0$ is given by the result obtained with the
OPE
\begin{equation}
 \rho^{\text{cont}}(s) = \rho^{\text{OPE}}(s)\theta(s-s_0),\label{eq:rho_cont}
\end{equation}
with $\theta(x)$ being the step function.
The OPE side is calculated up to leading order in $\alpha_s$ and
condensates up to dimension eight are considered. Currents and
OPE terms used in this work are taken from
Ref.~\cite{matheus07,Lee_Mihara_Navarra_Z4430_PLB,lee_dsdstar,bracco_z4430tetra,lee_morita_nielsen_molecule}.
The correlation function in the OPE side can be expressed as
\begin{equation}
 \Pi^{\text{OPE}}(q^2) = \int_{4m_c^2}^{\infty}ds
  \frac{\rho^{\text{OPE}}(s)}{s-q^2} + \Pi^{\text{mix}\langle \bar{q}q \rangle}(q^2),\label{eq:pi_ope}
\end{equation}
where $\rho^{\text{OPE}}(s) = \pi^{-1}\text{Im}\Pi^{\text{OPE}}(s)$.
Then, we can extract the pole term by equating the OPE expression
and the phenomenological expression, making the Borel transformation on
both sides and then transferring the continuum contribution
\eqref{eq:rho_cont} to the OPE side. The sum rule is hence given by
\begin{align}
 \int_{4m_c^2}^{\infty} ds \,e^{-s/M^2} \rho^{\text{pole}}(s) &=
  \int_{4m_c^2}^{s_0} ds \,e^{-s/M^2}\rho^{\text{OPE}}(s)\nonumber\\
 &+  \Pi^{\text{mix}\langle \bar{q}q \rangle}(M^2).\label{eq:sumrule}
\end{align}
Note that the left-hand side becomes $\lambda^2 e^{-m^2/M^2}$ in the
narrow width approximation. In this case, one can extract the pole mass by
taking the ratio between the derivative of Eq.~\eqref{eq:sumrule} with
respect to $1/M^2$ and  Eq.~\eqref{eq:sumrule} itself. In the present
work, we introduce the width by employing the Breit-Wigner function
to the pole contribution
\begin{equation}
 \rho^{\text{pole}}(s) = \frac{1}{\pi}\frac{f\Gamma
  \sqrt{s}}{(s-m^2)^2+s\Gamma^2}.\label{eq:BW}
\end{equation}

The mass and width are determined by looking at the stability of mass
against varying Borel mass $M^2$, as usual. The relevant Borel window is
 determined by the convergence of the OPE for the minimum
 $M_{\text{min}}^2$ and the pole dominance criterion for the maximum
 $M_{\text{max}}^2$. As usual, we determine $M_{\text{min}}^2$ by requiring 
the dimension eight condensate contribution to be less than 10\% of the total OPE and  $M_{\text{max}}^2$ from more than  50\% pole dominance.
We calculate the mass by
fixing a width and solving the equation for the ratio
\begin{equation}
 -\frac{1}{\Pi^{\text{OPE}}(M^2)}\frac{\partial
  \Pi^{\text{OPE}}(M^2)}{\partial (1/M^2)} =
  \frac{\int_{4m_c^2}^{\infty}ds
  se^{-s/M^2}\rho^{\text{pole}}(s)}{\int_{4m_c^2}^{\infty} ds e^{-s/M^2}\rho^{\text{pole}}(s)}
\label{eq:sumrule2}
\end{equation}
where $\Pi^{\text{OPE}}(M^2)$ is the right-hand side of
 Eq.~\eqref{eq:sumrule} and $\rho^{\text{pole}}(s)$ is given by
 Eq.~\eqref{eq:BW}.
\begin{figure}[!ht]
 \includegraphics[width=3.375in]{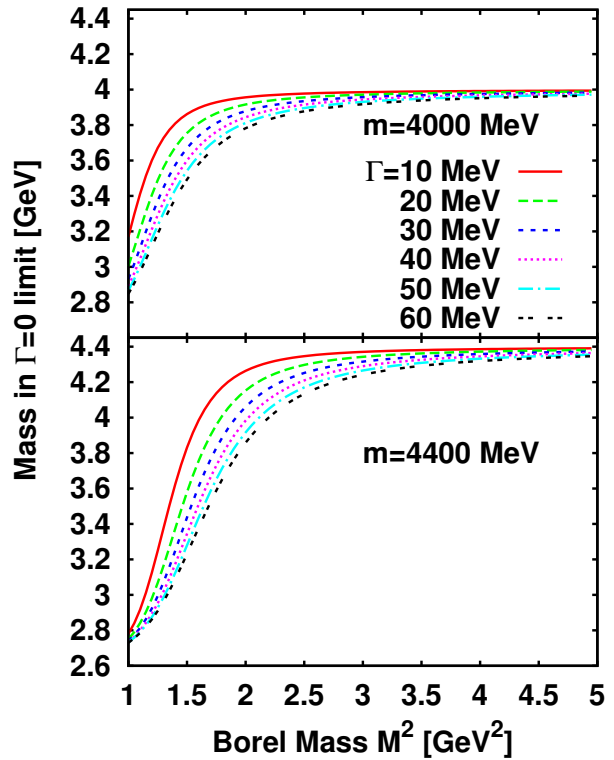}
 \caption{(color online). Right-hand side of Eq.~\eqref{eq:sumrule2} as a function of
 Borel mass $M^2$ with fixed $m$ and $\Gamma$. Upper and lower panels
 stand for the case of $m=4$ GeV and $m=4.4$ GeV, respectively.}
 \label{fig:mstar-m2_m}
\end{figure}
\begin{figure}[!ht]
  \includegraphics[width=3.375in]{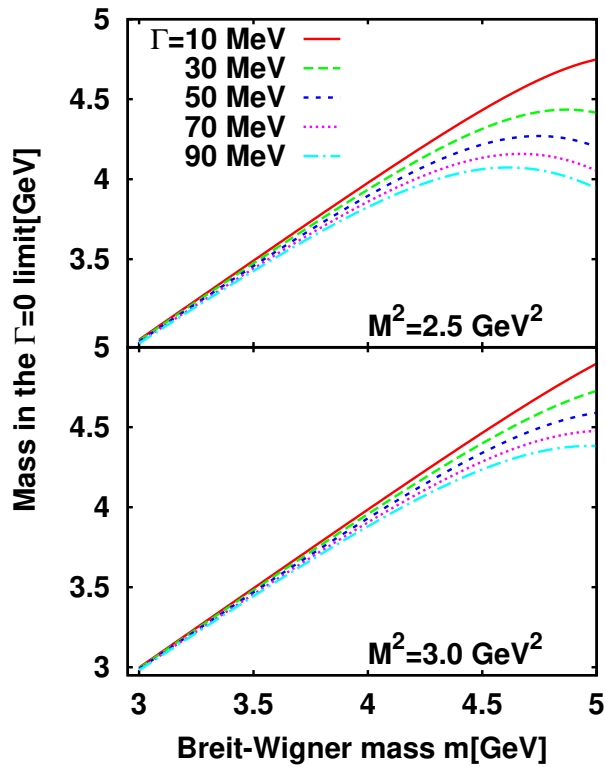}
 \caption{(color online.) Same as Fig.~\ref{fig:mstar-m2_m}, but as a
 function of the Breit-Wigner mass and for a fixed Borel mass.}
 \label{fig:bw_m2fix}
\end{figure}
 \section{General features}
\label{sec:width_general}

Since the left-hand side in Eq.~\eqref{eq:sumrule2} does not change by
introducing a width, one can investigate how the mass changes only by looking at
the behavior of the right-hand side.

Figure \ref{fig:mstar-m2_m} shows the right-hand side of
Eq.~\eqref{eq:sumrule2}  as a function of Borel mass $M^2$.
One can see that it is a monotonic function of $M^2$ if both $m$ and
$\Gamma$ are fixed. It rapidly increases at small $M^2$ and then
asymptotically reaches to the Breit-Wigner mass $m$. In analyses of
QCDSR, we solve Eq.~\eqref{eq:sumrule2} with a value of left-hand
side of Eq.~\eqref{eq:sumrule2} given by the OPE side and the continuum. This
procedure corresponds to finding an intersection between a horizontal line
denoting the value of mass in the $\Gamma=0$
and the curves of a given Borel mass in the figures.
For example, if one has $m = 3.8$ GeV in the $\Gamma=0$ case, a
possible solution at $M^2 = 2.0$ GeV$^2$ and $m=4$ GeV is
$\Gamma \simeq 40$ MeV. If one sets $m=4.4$ GeV, $\Gamma \simeq 60$ MeV
is one of the possible solutions. The best solution is determined by
looking at the stability against $M^2$.
From the monotonic behavior seen in Fig.~\ref{fig:mstar-m2_m}, one notes
that introducing the width increases the mass especially at small Borel
mass region. Hence, if one gets the mass in the $\Gamma = 0$
case which
monotonically increases as $M^2$ increases, it will be improved by
including the width. This fact gives a guideline on the QCDSR
analyses.\footnote{This is not a completely general result of QCDSR; as seen in
\cite{lee_morita_ohnishi_sigma}, introducing width leads to smaller mass 
when $M^2$ is large compared to $m$.}

We also plot the direct relation between the Breit-Wigner mass and
the mass in the $\Gamma=0$ case in Fig.~\ref{fig:bw_m2fix}. Here we
fixed the Borel mass
$M^2=2.5$ GeV$^2$ in the top panel and 3.0 GeV$^2$ in the bottom panel,
which are typical values satisfying the stability criterion in the QCDSR
analyses below. One can see that deviation from the mass in the
$\Gamma=0$ is larger for larger mass $m$ and smaller Borel mass
$M^2$. One should note that it is no longer monotonic
as a function of $m$ at smaller $M^2$ and large $\Gamma$, as seen in
the top panel.
This means that if one gets the mass 4200 MeV in the
$\Gamma = 0$ case with stability at $M^2 = 2.5$ GeV$^2$,
the maximum width of this state is limited to 50 MeV. This also
gives the constraint on possible mass and width values.
\begin{figure}[ht]
  \includegraphics[width=3.375in]{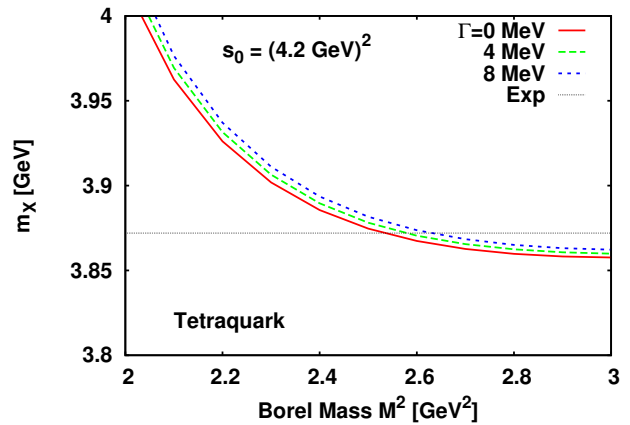}
  \caption{(color online.) Results for $X(3872)$ as a tetraquark state with
 width. Continuum thresholds $s_0$ is taken to be $\sqrt{s_0}=4.2$ GeV.}
  \label{fig:x3872_tetra}
\end{figure}
\begin{figure}[ht]
 \includegraphics[width=3.375in]{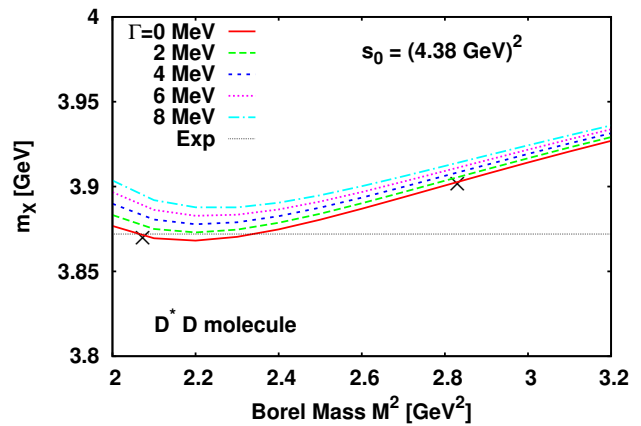}
 \caption{(color online). Results for $X(3872)$ as a $D^* D$ molecule.
 The crosses indicate lower and upper limit of the Borel window, respectively.}
 \label{fig:dstard}
\end{figure}
\section{Results}
\label{sec:results}

For parameters in the QCDSR analyses, we use the same parameter set as
in the previous works and
assume the factorization of the higher dimensional condensates.
Namely, $m_c=1.23$ GeV,
$\langle \bar{q}q \rangle = -(0.23 \text{GeV})^3$,
$\langle g^2 G^2 \rangle=0.88 \text{GeV}^4$,
$\langle \bar{q} g\sigma\cdot G q \rangle=m_0^2 \langle \bar{q}q\rangle$ and
$m_0^2=0.8 \text{GeV}^2$. Here we ignored the possible uncertainties in
these parameters. Because the uncertainties are those related to the OPE
side, which are unchanged by including the width, possible errors on masses
will not be so different from the ones estimated in the previous works.
 \begin{figure*}[ht]
 \includegraphics[width=6.75in]{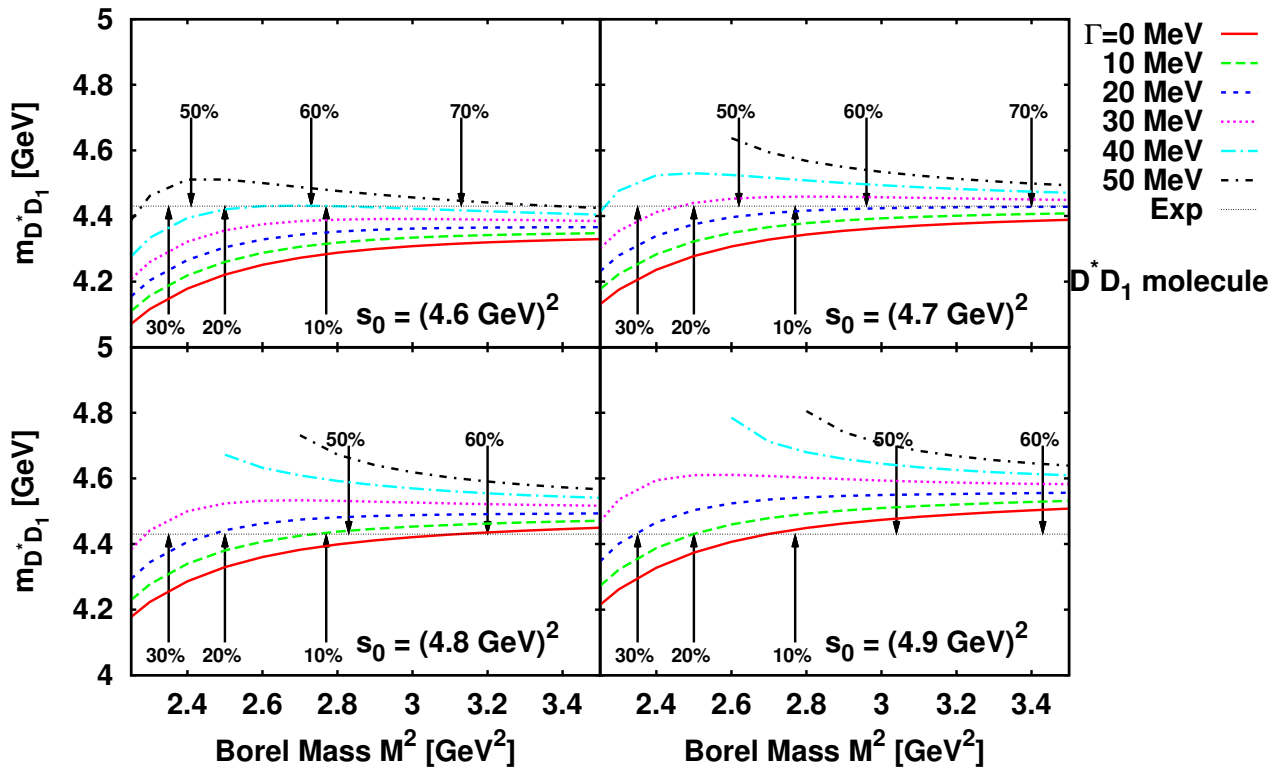}
  \caption{(color online). Results for $D^* D_1$ molecule state. Each panel shows
  different continuum threshold case. Upward and downward arrows
  indicate the region of the Borel window $M^2_{\text{min}}$ and
  $M^2_{\text{max}}$, respectively. Associated numbers in percent denote the
  dimension eight condensate  contribution for upward arrows and
  continuum contribution for downward ones.}
  \label{fig:DstarD1}
 \end{figure*}

From the consideration in the previous section, one finds that
tetraquark configurations for $Z(4430)$ examined in
Ref.~\cite{bracco_z4430tetra} are ruled out even if the width is taken
into account.
In the $J^P=0^-$ results (Fig.~3 of Ref.~\cite{bracco_z4430tetra}),
the mass in the $\Gamma=0$ case shows a good stability. Incorporating
width does not improve the stability, but it raises the value of mass,
which is already bigger than the experimental value in the $\Gamma=0$
case.  If one assumes $J^P=1^-$ tetraquark state, it is shown that
the mass is more than 300 MeV larger than the experimental value, and the functional
behavior with respect to $M^2$ is monotonically decreasing \cite{bracco_z4430tetra}.
Both of these features are only worsened by introducing the width in
the calculations. The failure to explain the $Z(4430)$ with
tetraquark structures cannot be corrected by taking width into account.
In the case of $X(3872)$, the value of mass in the $\Gamma=0$ limit is found
to be in agreement with the data \cite{matheus07} while
functional behavior against $M^2$ is monotonically decreasing.
In this case, however, the experimental width has been found to be around 2.3 MeV
\cite{choi03_belle_x3872}. Then, such a small width hardly improves the Borel curve, as shown in Fig.~\ref{fig:x3872_tetra}.\footnote{In
Ref.~\cite{matheus07}, the mass of $X(3872)$ was evaluated by including
condensates up to dimension five and higher dimensional ones were used
to estimate errors. In this paper, we have included all the condensates
given in Ref.\cite{matheus07}.}

Contrary to the sum rules obtained with interpolating currents based on large tetraquark component, molecular descriptions are much more
promising. First, let us consider a counter example of
Fig.~\ref{fig:x3872_tetra}.
The current for $D^* D$ molecule is
obtained by replacing the strange quark by a light quark in
Ref.~\cite{lee_dsdstar}:
\beq
j_\mu={i\over\sqrt{2}}\left[(\bar{u}_a\gamma_5 c_a)(\bar{c}_b\gamma_\mu
u_b)-(\bar{c}_a\gamma_5 u_a)(\bar{u}_b\gamma_\mu c_b)\right]\;,
\label{field}
\enq
where $a$ and $b$ are color indices. The combination
$D^{0}\bar{D}^{*0}-\bar{D}^0{D}^{*0}$ has $J^{PC}=1^{++}$ as the $X(3872)$ 
meson. To show that this current has positive
charge conjugation we notice that under charge conjugation transformation 
the two terms in Eq.~(\ref{field}) transforms as:
\beqa
\hat{C}[(\bar{u}_a\gamma_5 c_a)(\bar{c}_b\gamma_\mu u_b)]\hat{C}^{-1}&=&
-[(\bar{c}_a\gamma_5 u_a)(\bar{u}_b\gamma_\mu c_b)]
\nnb\\
\hat{C}[(\bar{c}_a\gamma_5 u_a)(\bar{u}_b\gamma_\mu c_b)]\hat{C}^{-1}&=&
-[(\bar{u}_a\gamma_5 c_a)(\bar{c}_b\gamma_\mu u_b)].
\enqa
Therefore, one obtains
\beq
\hat{C}j_\mu\hat{C}^{-1}=j_\mu.
\enq
The symmetrical combination, $D^{0}\bar{D}^{*0}+\bar{D}^0{D}^{*0}$, would
provide exactly the same mass, within our sum rule approach.

The resultant OPE expressions for $D^* D$ molecule are
obtained by  putting $m_s=0$ and $\langle\bar{s}s\rangle=\langle\bar{q}q
\rangle$ in the expressions in Ref.~\cite{lee_dsdstar}. 
Figure \ref{fig:dstard} shows the
result of mass of the $D^* D$ molecular state with $\Gamma=0,2,4,6,8$
MeV. One sees that the molecular description gives better stability
than the tetraquark case. Furthermore, although effect of width is not
larger, we can fit the experimental mass 3872 MeV and width
$\Gamma < 2.3$ MeV simultaneously with a continuum threshold
$\sqrt{s_0}=4.38$ GeV, and obtain a broad Borel window. Consequently,
it is strongly favored that $X(3872)$ is a $D^* D$ molecular state.
\begin{figure*}[ht]
  \includegraphics[width=6.75in]{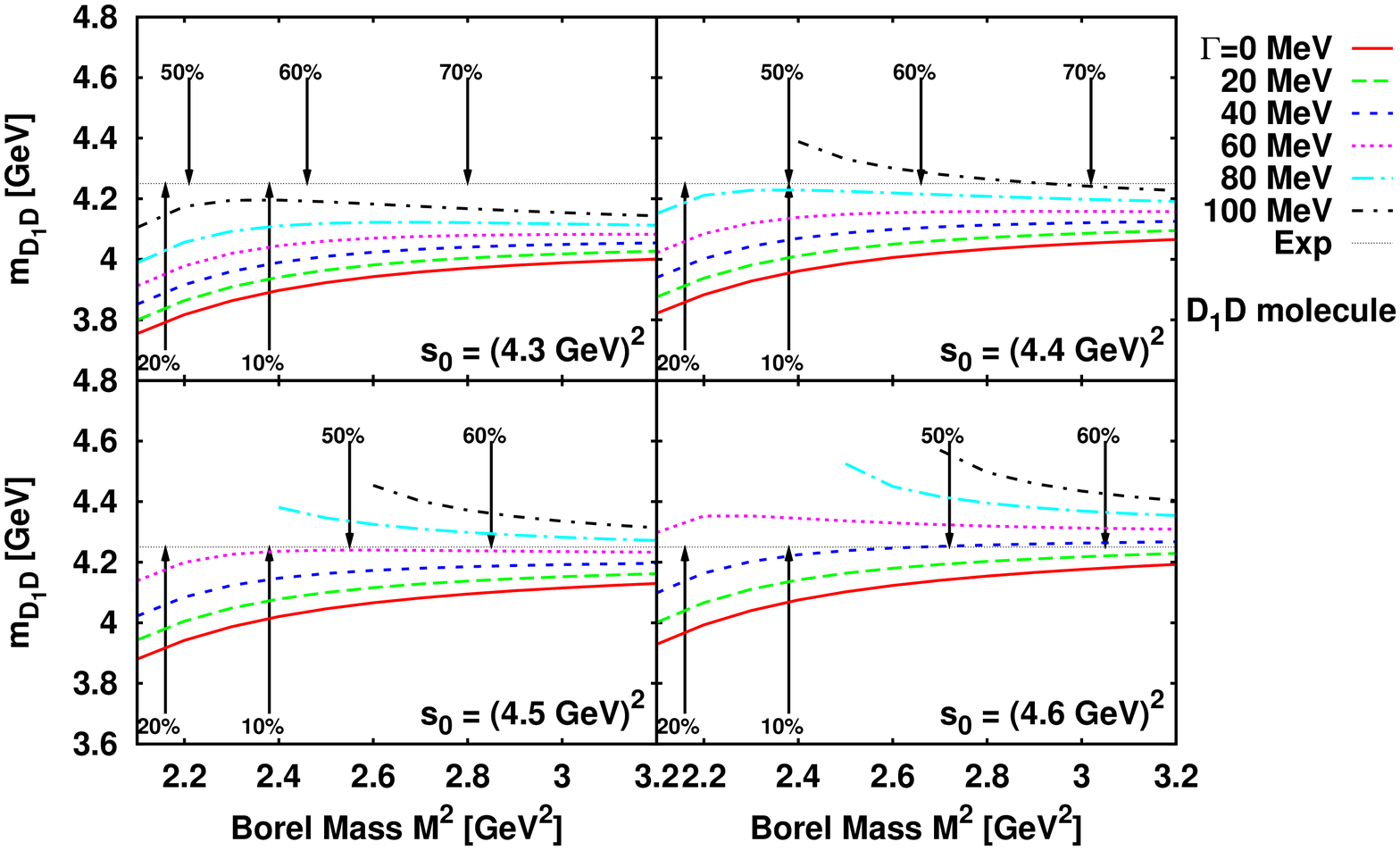}
  \caption{(color online). Results for $D_1 D$ molecule. Symbols are similar to
 Fig.~\ref{fig:DstarD1}.}
  \label{fig:d1d}
\end{figure*}

Next, we consider $Z(4430)$ states as a $D^* D_1$ molecule.
The current and OPE expressions are given in
Ref.~\cite{Lee_Mihara_Navarra_Z4430_PLB}. It has been shown that
the molecular description gives a mass which agrees with the
experiment well. Since the mass in the $\Gamma=0$ case is monotonically
increasing function of $M^2$, it is expected that incorporating the 
width improves the stability according to the result shown in
Sec.~\ref{sec:width_general}.

Figure \ref{fig:DstarD1} shows the result of $D^* D_1$ molecule with
various continuum thresholds. In
Ref.~\cite{Lee_Mihara_Navarra_Z4430_PLB}, the continuum threshold is
determined as $\sqrt{s_0}=4.8-5.0$ GeV. The center value
$\sqrt{s_0}=4.9$ GeV case is plotted in the right-bottom panel.
In this case, the mass in the $\Gamma=0$ case agrees well with the experimental
value.
As introducing the width raises the mass, however, the mass becomes larger
than experiment when $\Gamma$ is as large as the experiment. One notes
that the stability becomes much better when $\Gamma \simeq 30$ MeV.
Other three panels show the cases with lower continuum thresholds.
Especially $\sqrt{s_0}=4.6$ GeV case reproduces both mass and width
quite well. One can see that $\Gamma \simeq 40$ MeV gives the best
stability of the mass, which perfectly agrees with the the experiment.
In the lower continuum thresholds case, however, we have to relax the
criterion for the allowed region of sum rule analyses, \text{i.e.},
Borel window. Since the mass of $Z(4430)$ is close to $D^*D_1$
threshold, it might be plausible that the continuum contribution becomes larger
when $D^* D_1$ forms a molecule, as assumed in this calculation.
The arrows in the figure indicate the Borel masses determined from
various values of the relative contributions of the dimension eight
condensates (for $M_{\text{min}}^2$, upward arrows) and the continuum
contribution (for $M_{\text{max}}^2$, downwards arrows). One sees that
reasonable Borel windows open if one relaxes the condition for either
$M_{\text{min}}^2$ or $M_{\text{max}}^2$, or both of them.

One notes that there is a
truncation of the curves in each panel, especially for large width
data. This is due to the nature of the Breit-Wigner function, shown in
upper panel of Fig.~\ref{fig:bw_m2fix}, that the right-hand side of
Eq.~\eqref{eq:sumrule2} has the maximum at low Borel mass and large
width region.
This fact appears as an absence of the solution of Eq.~\ref{eq:sumrule2}
for a fixed Borel mass and $\Gamma$. Hence, it expresses
a maximum width allowed by the QCDSR for each value of the continuum
threshold.

Finally we consider the recently discovered
$Z_2^+(4250)$ as a $D_1 D$ molecule. The current and resultant OPE
expressions are given in Ref.~\cite{lee_morita_nielsen_molecule}.
In Ref.~\cite{lee_morita_nielsen_molecule}, it is shown that $D_1 D$
molecular description gives a reasonble agreement with the $Z_2^+(4250)$ mass.
As mentioned above, this state
has a  large width whose effect should be examined. Indeed, $M^2$
dependence of the mass looks promising because it is a monotonically
increasing function of $M^2$ as in the case of $D^*D_1$ molecule.

Figure \ref{fig:d1d} shows the results for the $D_1D$ molecule.
As in Fig.~\ref{fig:DstarD1}, we plot the mass for various widths
in each panel. The continuum threshold $\sqrt{s_0}=4.6$ GeV corresponds to
the center value in
the previous analysis \cite{lee_morita_nielsen_molecule}. One can see
that the stability is achieved in all the cases. In the largest
$\sqrt{s_0}$ case, the optimized width is around 40 MeV, which is a
little smaller than the experimental value. Reducing the continuum
threshold leads to larger width; for $\sqrt{s_0}=4.5$ GeV, we obtain the
$\Gamma=60$ MeV. Up to this value, no relaxation of the Borel window
criterion is needed. Taking $\sqrt{s_0} \leq 4.4$ GeV makes the width
much closer to the experiment, however, we need to relax the condition for
the dimension eight condensates and/or the continuum contribution  as in the case of $Z(4430)$ to validate QCDSR.
In such cases, we can obtain $\Gamma\simeq 80$ MeV ($\sqrt{s_0}=4.4$
GeV) and $\Gamma \simeq 100$ MeV ($\sqrt{s_0}=4.3$ GeV). Hence, our
analyses support an existence of $D_1 D$ molecular state with large
width and its manifestation as $Z_2^+(4250)$ resonance.

\section{Summary}
\label{sec:summary}

In summary, we have extended previous QCDSR analyses of exotics to include the
total width, by employing the Breit-Wigner function to the pole term. As
a general
feature, for the cases where the predicted mass for $\Gamma=0$ increases with increasing Borel mass $M^2$, introducing the width increases the predicted mass at small Borel mass region, and improves the Borel stability,
From this point of view, none of the sum rules based on interpolating currents with tetraquark components are favored.
On the other hand, sum rules based on interpolating currents with molecular description as $D^* D$, $D^* D_1$ and $D_1 D$, are shown to give valid sum rules for $X(3872)$, $Z^+(4430)$ and $Z_2^+(4250)$ respectively.  For $X(3872)$, the inclusion  of the
width slightly modify the mass, leading to a better agreement with the
experimental result. For $Z^+(4430)$ and $Z_2^+(4250)$, molecular
description proposed in
Refs.~\cite{Lee_Mihara_Navarra_Z4430_PLB,lee_morita_nielsen_molecule} are
largely improved by introducing the width. We have obtained stable results with
$\Gamma_{Z(4430)}\simeq 40$ MeV and $\Gamma_{Z_2(4250)}\simeq 40-100$
MeV. These results strongly support the previous results based on
$\Gamma \rightarrow 0$ limit, that the $Z^+(4430)$ and $Z_2^+(4250)$
resonances are strong candidates for molecular states.   Moreover, we have established that the QCD sum rule with four OPE terms has sufficiently rich structure so that an estimate of the total width is also possible.

\section*{Acknowledgment}
This work has been partly supported by BK21 program of the Korean
Ministry of Education, Korean Research Foundation KRF-2006-C00011, and
FAPESP and CNPq-Brazil.


\end{document}